# k-space imaging of anisotropic 2D electron gas in GaN/GaAlN high-electron-mobility transistor heterostructures

L. L. Lev,[1,2,*] I. O. Maiboroda,[2,*] M.-A. Husanu,[1,3] E. S. Grichuk,[2] N. K. Chumakov[2], I. S. Ezubchenko,[2] I. A. Chernykh[2], X. Wang,[1] B. Tobler,[1] T. Schmitt,[1] M. L. Zanaveskin,[2] V. G. Valeyev[2] & V. N. Strocov[1,*]

[1] Swiss Light Source, Paul Scherrer Institute, 5232 Villigen-PSI, Switzerland

[2] National Research Centre "Kurchatov Institute", 123182 Moscow, Russia

[3] National Institute of Materials Physics, Atomistilor 405A, 077125-Magurele, Romania

* These authors have contributed equally to this work. Correspondence and requests for materials should be addressed to V.N.S. (email: vladimir.strocov@psi.ch)

**Nanostructures based on buried interfaces and heterostructures are at the heart of modern semiconductor electronics as well as future devices utilizing spintronics, multiferroics, topological effects and other novel operational principles. Knowledge of electronic structure of these systems resolved in electron momentum k delivers unprecedented insights into their physics. Here, we explore 2D electron gas formed in GaN/AlGaN high-electron-mobility transistor (HEMT) heterostructures with an ultrathin barrier layer, key elements in current high-frequency and high-power electronics. Its electronic structure is accessed with angle-resolved photoelectron spectroscopy whose probing depth is pushed to a few nm using soft-X-ray synchrotron radiation. The experiment yields direct k-space images of the electronic structure fundamentals of this system – the Fermi surface, band dispersions and occupancy, and the Fourier composition of wavefunctions encoded in the k-dependent photoemission intensity. We discover significant planar anisotropy of the electron Fermi surface and effective mass connected with relaxation of the interfacial atomic positions, which translates into non-linear (high-field) transport properties of the GaN/AlGaN heterostructures as an anisotropy of the saturation drift velocity of the 2D electrons.**

The concept of HEMTs advanced by T. Mimura in the early 80s[1] for GaAs/GaAlAs heterostructures revolutionized the field of high-frequency semiconductor electronics. It exploits an idea of polarization engineering when a large band offset between an intrinsic semiconductor and a doped barrier layer forms a quantum well (QW) at the interface that confines a mobile 2D electron gas (2DEG) on the intrinsic semiconductor side. Its spatial separation from defects in the doped barrier layer and in the interface region - in contrast to conventional transistor structures where the 2DEG is formed by doping - allows the electrons to escape defect scattering and dramatically increase their mobility $\mu_e$, limited then only by phonon scattering. This fundamental operational principle of HEMTs boosts their high-frequency performance, which is exploited in a wide spectrum of applications, such as cell phones.

A characteristic property of HEMTs based on wurtzite GaN/AlGaN heterostructures is the accumulation of large sheet carrier concentrations $n_s \sim 10^{13}$ cm$^{-2}$ - about one order of magnitude higher compared to other III-V or elementary semiconductors[2,3] - without intentional doping of the barrier layer. This property is attributed to the formation of a deep spike-shaped QW at the heterojunction, where a large conduction-band offset coexists with large piezoelectric and spontaneous polarization[4,5]. Although $\mu_e$ in GaN-HEMTs is limited by a relatively large electron effective mass $m^* \sim 0.22$ in bulk GaN, nearly three times larger than in GaAs, these devices demonstrate an advantageous combination of sufficiently high operating frequency with exceptionally high current density, resulting from the large $n_s$, saturation drift velocity, operating temperature and breakdown voltage. These advantages make the GaN-HEMTs indispensable components of high power amplifiers for microwave communication and radar systems. Recently, the ideas of creating mobile 2DEGs using spontaneous and strain-induced polarization at the interface have been extended to oxide systems such as the binary $Mg_xZn_{1-x}O/ZnO$,[6] $LaAlO_3/SrTiO_3$,[7] and $CaZrO_3/SrTiO_3$[8] heterostructures that typically embed orders of magnitude larger $n_s$.

The state-of-art GaN-HEMTs operate nowadays at the edge of their physical limitations, which remain far from complete understanding. Development of new strategies to improve their performance and conquer the near-THz operational range[9] needs qualitatively new experimental knowledge about the physics of these devices. Particularly important is **k**-space information about the Fermi surface (FS), band dispersions and wavefunctions of the embedded 2DEG. These fundamental electronic structure characteristics, only indirectly accessible in optics and magnetotransport experiments such as the Hall effect, Shubnikov-de Haas oscillations, cyclotron resonance, etc.[10,11,12], can be directly accessed probed using the **k**-resolving technique of angle-resolved photoelectron spectroscopy (ARPES). However, the small photoelectron mean free path $\lambda_{PE}$ in conventional ARPES with photon energies $h\nu$ around 100 eV limits its depth sensitivity to ~0.5 nm. Access to buried electron systems such as the HEMTs requires pushing this technique to the soft-X-ray energy range (SX-ARPES, see a recent review Ref. 13) with $h\nu$ around 1 keV and higher, where $\lambda_{PE}$ grows with photoelectron kinetic energy as ~$E_k^{3/4}$ [14,15]. For GaN in particular, elastic-peak electron spectroscopy measurements[16] show that an increase of $E_k$ from 200 eV to 2 keV results in an increase of $\lambda_{PE}$ from ~0.5 to 4 nm. An added virtue of SX-ARPES, still largely overlooked in applications to two-dimensional (2D) systems like quantum-well states (QWSs), is a significantly sharper definition of the out-of-plane component $K_z$ of the final-state momentum **K**. This fact results from the larger $\lambda_{PE}$ translating, via the Heisenberg uncertainty principle, to a sharper definition of $K_z$[17,18]. As we will see below, in this case the ARPES signal provides the Fourier composition of the 2DEG wavefunctions. SX-ARPES on buried systems, challenged by photoelectron attenuation in the overlayers as well as a progressive reduction of photoexcitation cross-section of the valence band (VB) states with $h\nu$[19], requires advanced synchrotron radiation sources delivering high photon flux (see Methods).

On the sample fabrication side, the 2DEG in GaN-HEMT heterostructures has until recently remained inaccessible to SX-ARPES because of the prohibitively large – of the order of 20-30 nm – depth of the AlGaN barrier layers. However, recent progress in MBE technology, in pursuit of yet higher operation frequencies of these devices, has allowed fabrication of heterostructures with ultrathin barrier layers of 3-4 nm[20,21,22] which make them ideally suited to SX-ARPES. This has allowed direct **k**-space imaging of the fundamental electronic structure characteristics – the Fermi surface, electron dispersions and the Fourier composition of wavefunctions – of the interfacial 2DEG in such heterostructures.

## Results

**Fabrication and basic electronic properties of GaN-HEMT heterostructures.** Our samples were grown on *c*-oriented sapphire substrate. The 500 nm thick Ga-polar GaN layer was grown on top of an AlGaN buffer layer required to suppress the crystal defects and promote growth of a smooth uniform film[23]. GaN layer was overgrown by a barrier layer consisting of 2 nm of AlN and 1 nm of $Al_{0.5}Ga_{0.5}N$, see Fig. 1a. Hall-effect measurements on our samples, 1b, have found $n_s$ ~ $8.2 \cdot 10^{12}$ cm$^{-2}$ almost constant through the temperature range $T = 5 \div 300K$ that confirms the high quality of the fabricated structures.

A simplified electronic structure model of our GaN-HEMTs is shown in Fig. 1b. It was evaluated within the conventional envelope-function approach (neglecting atomic corrugation) by self-consistent solutions of the 1D Poisson-Schrödinger equations with the Dirichlet boundary conditions adjusted to reproduce the experimental $n_s$ (for details see Supplementary Note 1). The effective 1D interfacial potential $V(z)$ as a function of out-of-plane coordinate $z$ confines two QWSs with different spatial localization. The QWS$_1$ embeds larger partial $n_s$ and is localized closer to the interface compared to the QWS$_2$, which is shifted into the $V(z)$ saturation region.

**Fermi surface: Anisotropy and carrier concentration.** A scheme of our SX-ARPES experiment is presented in Fig. 1a. The data were acquired in the $hv$ region between 800 and 1300 eV where the interplay of photoelectron transmission through the barrier overlayer, increasing with $hv$, and photoexcitation cross-section, decreasing with $hv$, maximizes the QWS signal. In view of the even QWS wavefunction symmetry, we used $p$-polarization of the incident X-rays, which minimizes the geometry- and polarization-related matrix element effects which distort the direct relation between ARPES intensities and Fourier composition of the QWS wavefunctions (see below).

The experimental FS map measured as a function of in-plane momentum $\mathbf{K}_{xy} = (K_x, K_y)$ at $hv = 1057$ eV (maximizing the QWS signal, see below) is presented in Fig. 2a. The FS formed by the QWSs in the HEMT channel appears as tiny circles whose $\mathbf{k}_{//}$ is located around the $\bar{\Gamma}$-points of the heterostructure's 2D Brillouin zone (BZ) shown in Fig. 1c. The location of the FS pockets coincides with the VB maxima (VBM), as seen in an iso-$E_B$ map of the VB shown in Fig. 2b. In the direct band gap GaN, this location in $\mathbf{k}_{//}$ is consistent with the conduction band minimum (CBM) derived character of the QWSs.

A high-resolution cut of the FS in Fig. 2c displays an external contour with a diameter of ~0.15 Å$^{-1}$ manifesting the QWS$_1$ and finite spectral weight in the middle, suggesting the presence of QWS$_2$. The latter is confirmed by the momentum-distribution curves (MDCs) of the Fermi intensity around the $\bar{\Gamma}_{10}$-points (Fig. 2d) where a weak structure in the middle signals the QWS$_2$. Our observation of QWS$_2$ is consistent with its recent detection via magneto-transport spectroscopy[24]. The presence of this state centered further away from the interface compared to the QWS$_1$ indicates that $V(z)$ in the GaN-HEMTs has a long-range saturated shape (see the model in Fig. 1b). The external Fermi intensity peaks in the $E_F$-MDCs directly identify the Fermi momenta $k_F$ of the QWS$_1$.

Such a comprehensive $\mathbf{k}$-space view of the buried QWSs, achieved with SX-ARPES, delivers two important observations. First, the experimental $E_F$-MDCs reveal significant anisotropy of the FS characterized by a difference of the $k_F$ values of 0.095±0.006 Å$^{-1}$ along the $\overline{\Gamma M}$ azimuth and 0.085±0.004 Å$^{-1}$ along $\overline{\Gamma K}$. These values were determined from the maximal gradient of the Fermi intensity[25] which delivers accurate $k_F$-values even when the bandwidth is comparable with the experimental resolution. The indicated statistical errors are the standard deviations of the $k_F$ values over the measurement series, and scaled with the Student's *t*-distribution coefficients for a confidence interval of 76%. The corresponding $k_F$ anisotropy factor $A_F = \frac{k_F^{\overline{\Gamma K}} - k_F^{\overline{\Gamma M}}}{\langle k_F \rangle}$, where $\langle k_F \rangle = (k_F^{\overline{\Gamma K}} + k_F^{\overline{\Gamma M}})/2$, measures in our case ~11%. Such a considerable FS anisotropy has been completely overlooked in previous macroscopic experimental studies without **k**-resolution.

Why does the 2DEG show such a pronounced planar anisotropy? Our simulations of electronic structure of the GaN-HEMTs (see Methods) used the standard density-functional theory (DFT) which is known to well describe excitation energies in GaN apart from the electron exchange-correlation discontinuity across the band gap approximated by the so-called scissor operator[26]. As the QWSs inherit their wavefunctions from the CBM of the bulk GaN (see below) we have started our analysis from the bulk band structure (shown later in Fig. 4). The corresponding $A_F$ in Fig. 2e is however negligible through the whole range of <$k_F$> defining the band filling. Next, we approached the electronic structure of our heterostructure system using GaN/AlN slab calculations with the unit cell (u.c.) shown in Fig. 2g. The atomic positions were relaxed under the constraint of the bulk GaN lateral lattice constants and symmetry. The resulting *c*-oriented Ga-N bond length $d_v$ in Fig. 2f shows a significant increase towards the interface relative to the bulk value. The corresponding atomic displacement contributes to the piezoelectric polarization at the GaN/AlN interface. The layer-resolved electronic structure of this system was characterized by the **k**$_{//}$-resolved layer density of states (LDOS) defined as, $\rho_z(\mathbf{k}_{//}, E) = \int_\Omega dx dy \sum_n |\psi_n(\mathbf{r}, \mathbf{k}_{//}, E)|^2$, where **r**=(x,y,z), the summation includes all *n*-th electron states with wavefunctions $\psi_n$ available for given **k**$_{//}$ and *E*, and the integration runs over the lateral unit cell Ω. Fig. 2h shows the LDOS calculated near the CBM (the VB results are given in Supplementary Note 2) for the interface, sub-interface and deep bulk-like GaN layers, where the bottom of the LDOS continuum corresponds to the QWS$_1$. The corresponding $A_F$ plots in 2e now show significant anisotropy, increasing towards the interface. At the experimental <$k_F$> the interface layer $A_F$ is ~7% which falls almost within the error bars of the experimental value. Finally, we performed the same LDOS calculations with the atomic coordinates in the slab fixed at the bulk values (without relaxation). $A_F$ immediately returned to the negligible bulk values. This analysis suggests thus that the discovered 2DEG anisotropy in GaN-HEMTs is a purely interface effect caused by relaxation of atomic position near the GaN/AlN interface.

We note however that the predicted atomic relaxation is restricted to a few atomic layers next to the interface, and it is not clear why it should significantly affect the 2DEG, whose maximal density is located ~3 nm away from this region (Fig. 1b). Actually, even the state-of-art growth methods leave significant intermixing of Ga and Al atoms at the GaN/AlN interface[21], resulting in a gradual variation of the lattice parameters over 1-3 nm from the interface. In the spirit of the entanglement between atomic relaxation and LDOS anisotropy, revealed by our computational analysis, this lattice distortion may cause significant electronic structure anisotropy extending into the 2DEG localization region.

Another important characteristic of the buried 2DEG is the experimental FS area which, by the Luttinger theorem[27], directly related to the $n_s$ sheet carrier concentration. In our case, the area of the external QWS$_1$ contour translates into partial $n_s^{(1)}$ = (12.8±1.4)·10$^{12}$ cm$^{-2}$, and that of the internal QWS$_2$ contour into $n_s^{(2)}$ = (0.5±0.4)·10$^{12}$ cm$^{-2}$. We note that the emergence of two QWSs goes together with large $n_s$ formed in the anomalously deep QW of the GaN-HEMTs. In our case the QWS$_2$ contributes only ~4% of the total $n_s$ dominated by the QWS$_1$. A significant difference of this ratio to that of 17% found in a cyclotron resonance study[28] is explained by extremely high sensitivity of the QWS$_2$ population to the interfacial QW depth in different samples. The total $n_s$ in our case amounts to (13.3±1.8)·10$^{12}$ cm$^{-2}$ which is in fair agreement with $n_s$ = 8.2·10$^{12}$ cm$^{-2}$ obtained by our Hall characterization, see Fig. 1d, in particular taking into account a small systematic overestimate of $k_F$ introduced by the gradient method[25].

**Momentum dependence of ARPES intensity: Wavefunction character.** Within the one-step theory of photoemission – see, for example, ref. 29 – the ARPES intensity is found as $I_{PE} \propto |\langle f|\mathbf{A}\cdot\mathbf{p}|i\rangle|^2$, where $\langle f|$ is the final and $|i\rangle$ the initial states coupled through the vector potential $\mathbf{A}$ of the incident electromagnetic field and momentum operator $\mathbf{p}$. Neglecting the experimental geometry and polarization effects, this expression simplifies to the scalar product $I_{PE} \propto |\langle f|i\rangle|^2$. For sufficiently high photon energies, $\langle f|$ approximates a plane wave $e^{i\mathbf{Kr}}$ periodic in the in-plane $xy$-direction and damped in the out-of-plane direction z, and the ARPES intensity appears [30,31] as $I_{PE} \propto |\langle e^{i\mathbf{Kr}}|i\rangle|^2$. We will now apply this formalism to the QWS wavefunctions $\psi_{QWS}$.

We will first analyze $I_{PE}$ as a function of photoelectron $\mathbf{K}_{xy}$ in-plane momentum. If we represent $\psi_{QWS}$ by Fourier expansion over 2D reciprocal vectors $\mathbf{g}$ as $\psi_{QWS}(\mathbf{r}) = \sum_{\mathbf{g}} A_{\mathbf{k}_{xy}+\mathbf{g}} e^{i(\mathbf{k}_{xy}+\mathbf{g})\mathbf{r}_{xy}}$, the plane-wave orthogonality will select from the sum only the component whose in-plane momentum $\mathbf{k}_{xy}+\mathbf{g}$ matches $\mathbf{K}_{xy}$ (corrected for the photon momentum $p = h\nu/c$), i.e. $I_{PE}(\mathbf{K}_{xy}) \propto |A_{\mathbf{K}_{xy}}|^2$. Therefore, the $(K_x,K_y)$-dependent ARPES maps in Fig. 2a-c visualize essentially the 2D Fourier expansion of $\psi_{QWS}$.[30,31]

We will now analyze $I_{PE}$ as a function of photoelectron $K_z$ out-of-plane momentum varied in the experiment through $h\nu$. The corresponding iso-$E_B$ map in $(K_x,K_z)$ coordinates near the VBM is displayed in Fig. 3a. Its $K_z$-dispersive contours demonstrate the three-dimensional (3D) character of the VB states inherited from bulk GaN. The FS in Fig. 3b formed by the QWSs demonstrates a different behavior however: the ARPES signal sharply increases whenever $K_z$ approaches values of integer $G_z$ - corresponding to the Γ-points of the bulk BZ, but without any sign of $K_z$-dispersion. The latter is emphasized by the zooms-in Fig. 3d,e where the QWSs form segments straight in the $K_z$-direction. This pattern is characteristic of the 2D nature of the QWSs. The corresponding Fermi intensity is represented by $E_F$-MDCs in Fig. 3c that show periodic oscillations peaked where $K_z$ matches the Γ-points.

Why do the QWSs display such an ARPES response, periodically oscillating as a function of $h\nu$? The dependence of $\psi_{QWS}$ on $z$ for given $\mathbf{K}_{xy}$ is represented as $\psi_{QWS}(z) = E(z) \cdot B_{k_z}(z)$, where the envelope function $E(z)$ confines the oscillating Bloch wave $B_{k_z}(z)$ whose $k_z$ momentum is adapted to the 3D crystal potential[32]. In our case the first term is an Airy-like function embedded in the approximately triangular $V(z)$, and the second one derives from bulk states of the GaN host. Inheriting ideas of early ARPES studies on surface states[32,33], it can be shown[34] that if we expand $B_{k_z}(z)$ over out-of-plane reciprocal vectors $G_z$ of the 3D host lattice as $B_{k_z}(z) = \sum_{G_z} C_{G_z} e^{i(k_z+G_z)z}$, then the $K_z$-dependence of ARPES intensity for given $\mathbf{K}_{xy}$ appears as a sequence $I_{PE}(K_z) \propto \sum_{G_z} |C_{G_z}|^2 P(K_z - (k_z + G_z))$ of peaks $P(K_z - (k_z + G_z))$ centered at $K_z = k_z + G_z$. Physically, the ARPES intensity blows up whenever the photoelectron $K_z$ hits a $k_z+G_z$ harmonic of $\psi_{QWS}$ to maximize the $\langle f|$ and $|i\rangle$ scalar product. Amplitudes of the $I_{PE}(K_z)$ peaks $\propto |C_{G_z}|^2$ image Fourier composition of the oscillating $B_{k_z}(z)$ term of $\psi_{QWS}$ (modulated by $h\nu$-dependent photoelectron transmission through the AlGaN/AlN layer) and the peak shapes are related to the Fourier transform of the $E(z)$ term weighted by $e^{-\lambda_{PE}z}$.[34]

Importantly, the experimental $K_z$-dependence of the QWS signal in Fig. 3b exhibits peaks exactly at $k_z+G_z$, corresponding to the Γ-point of bulk GaN. In combination with the $(K_x,K_y)$-image in Fig. 2 where the QWS signal corresponds to the same Γ-point, this fact confirms that the $\psi_{QWS}$'s are derived from the CBM states of bulk GaN. In a methodological perspective, such identification of the $\psi_{QWS}$ character can be essential, for example, for heterostructures of layered transition metal dichalcogenides, where the CB can include two or more valleys almost degenerate in energy but separated in **k**. The knowledge of the QWS character will then allow the predictive manipulation of the valley degree of freedom for potential valleytronics devices[35].

We note that the common models of QWSs based on the 1D potential $V(z)$, like that in Fig. 1c, imply that their in-plane behavior $\psi_{QWS}(\mathbf{r}_{xy})$ is described by one single plane wave $e^{i\mathbf{K}_{xy}\mathbf{r}_{xy}}$ (i.e. one non-zero $A_{\mathbf{K}_{xy}}$ component), and out-of-plane behavior $\psi_{QWS}(z)$ is identical to the smooth $E(z)$ function. However, the experimental FS maps in Fig. 2a,c reveal numerous non-zero $A_{\mathbf{K}_{xy}}$ spread through **k**-space, and the $I_{PE}(K_z)$ oscillations numerous $k_z+G_z$ harmonics. Accurate QWS models should therefore incorporate atomic corrugation of the interfacial potential in the in-plane and out-of-plane directions to form $\psi_{QWS}$ as a confined Bloch wave.

The experimental distribution of high-energy ARPES intensity over sufficiently large volume of the $(K_x,K_y,K_z)$-space will in principle allow, notwithstanding the phase problem, a full reconstruction of $\Phi_{QWS}(x,y,z)$ in all three spatial coordinates, similar to the reconstruction of molecular orbitals (see Refs. 30,31 and references therein). This reconstruction will naturally incorporate full $\psi_{QWS}$ including the envelope and Bloch wave terms that goes beyond the common 1D models like in Fig. 1c describing the QWSs as free 2D electrons with empirical $m^*$ confined in the $z$-direction. More accurate models of the GaN-HEMTs should replace free electrons by Bloch ones, naturally incorporating atomic corrugation.

**Band dispersions: Effective mass.** Experimental band dispersions in GaN-HEMTs shown in Fig. 4 were measured along $\overline{\Gamma M}$ (a) and $\overline{\Gamma K}$ (b) at $hv = 1066$ eV bringing $K_z$ to the Γ-point of the bulk BZ. Non-dispersive ARPES intensity coming from the AlN and AlGaN layers is suppressed in these plots by subtracting the angle-integrated spectral component. The CBM-derived QWSs appear as tiny electron pockets above the VB dispersions of GaN. Their energy separation from the VBM is consistent with the GaN fundamental band gap of ~3.3 eV. Whereas the VB dispersions are broadened in $E_B$ primarily because of band bending in the QW region, the QWS dispersions stay sharp. This confirms their 2D nature insensitive to band bending as well as their localization in the deep defect-free region in GaN, spatially separated from the defect-rich GaN/AlN interface region, the fundamental operational principle of the HEMTs delivering high $\mu_e$. The panels Fig. 4c,f,i show the experimental dispersions as a function of $k_z$. Whereas clear dispersion of the VB states manifests their 3D character, the QWS are flat in $K_z$.

A zoom-in of the QWS dispersions along the $\overline{\Gamma M}$ and $\overline{\Gamma K}$ azimuths is shown in Fig. 4d-f with the corresponding MDC in Fig. 4g-i. Whereas the outer contour of these dispersions corresponds to the QWS$_1$, the significant spectral weight in the middle is due to the QWS$_2$. Following the $k_F$ anisotropy discussed above, a parabolic fit of the QWS$_1$ dispersions yields $m^*$ values of $(0.16 \pm 0.03)m_0$ along the $\overline{\Gamma M}$ azimuths and $(0.13 \pm 0.02)m_0$ along $\overline{\Gamma K}$ ($m_0$ is the free-electron mass) which thus differ from each other by ~22%.

Our SX-ARPES experiment presents thus a direct evidence of the planar FS and $m^*$ anisotropy in GaN-HEMTs. This effect was overlooked in previous studies because the optics methods are **k**-integrating, and quantum oscillations techniques lose their **k**-resolution in the 2D case. Magnetotransport experiments give only an indirect information on the 2DEG's $m^*$ [10,11,12] which is conventionally[36,37] assumed to be isotropic. We conjecture that further progress of SX-ARPES on energy resolution will push data analysis from merely band structure to one-electron spectral function $A(\omega,\mathbf{k})$ which will inform, for example, the interaction of electrons with other quasiparticles such as the phonon-plasmon coupled modes.[38]

Implications for the transport properties. How will the observed lateral anisotropy of the 2DEG electronic structure affect the transport properties? Naively, one might think that it would directly translate into an anisotropy of the electrical conductivity. However, fundamental linear response considerations attest that any physical properties such as conductivity described by a second-order tensor with C6 symmetry must be scalar, i.e. in the linear (low-field) regime, conductivity in the hexagonal lattice of GaN must be isotropic (Supplementary Note 3). On the other hand, this restriction is lifted for the non-linear (high-field) regime where conductivity can become anisotropic. A canonical example of such a crossover is n-doped Ge.[39,40] Although its FS is anisotropic, cubic symmetry of the Ge lattice results in isotropic low-field conductivity. However, with an increase of the electric field, conductivity along the ⟨**001**⟩, ⟨**011**⟩ and ⟨**111**⟩ crystallographic directions develops differently. The GaN-HEMTs can be easily driven into the non-linear regime where electronic current saturates due to electron scattering on longitudinal optical (LO) phonons[41,42]. To reach the LO phonon energy, lighter electrons should gain a larger drift velocity $V_{sat}$. Therefore, larger $V_{sat}$ and thus saturation current $I_{sat}$ should be expected in the directions of lower $m^*$.

We have examined low- and high-field conductivity in our GaN-HEMT heterostructures using samples essentially identical to the ARPES ones, but with the $Al_{0.45}Ga_{0.55}N$ layer thickness increased to 15 nm to prevent a 2DEG degradation during longer sample handling in air. Hall measurements showed $n_s = 2*10^{13}$ cm$^{-2}$, $\mu_e = 1150$ cm$^2$*V$^{-1}$*s and sheet resistance $R_s = 240$ Ohm*sq$^{-1}$ for these samples. The fabricated test modules were oriented at four different angles (0°, 30°, 60°, 90°) with respect to the substrate to promote current flow along the $\overline{\Gamma M}$ and $\overline{\Gamma K}$ azimuths (Fig. 5a,b). Results of the transport measurements presented in Fig. 5c show, as expected, isotropic low-field $R_s$. However, the $I_{sat}$ characteristic of the high-field regime is clearly anisotropic: 1.53±0.01 A/mm along $\overline{\Gamma K}$ and 1.46±0.01 A/mm along $\overline{\Gamma M}$ (Fig 5c). As a consistency check, the modules rotated by 60° with respect to each other showed the same $I_{sat}$ values, in accordance with hexagonal symmetry of the GaN electronic structure. These results on the previously overlooked $I_{sat}$ anisotropy demonstrate that $m^*$ along $\overline{\Gamma K}$ is smaller compared to $\overline{\Gamma M}$, as predicted by our ARPES results.

## Discussion

Our direct **k**-space imaging of the fundamental electronic structure characteristics – FS, band dispersions and occupancy, Fourier composition of wavefunctions – of the 2DEG formed in high-frequency GaN-HEMTs with ultrathin barrier layer makes a quantitative step compared to conventional optics and magnetotransport experiments. We discover, in particular, significant planar anisotropy of the 2DEG band dispersions caused by piezoelectrically active relaxation of atomic position near the GaN/AlN interface. This effect is found to manifest itself in non-linear electron transport properties as anisotropy of the saturation drift velocity and current. Our findings suggest a positive use of the crystallographic orientation to improve these high-power characteristics of GaN-HEMTs. Furthermore, our **k**-space image of the Fourier composition of the 2DEG wavefunctions calls for extension of the conventional 1D models of semiconductor interfaces to 3D ones based on the Bloch-wave description naturally incorporating atomic corrugation. The fundamental knowledge about GaN-HEMTs achieved in our work as well as new device simulation methods can clarify the physical limits of these heterostructures and finally push their reliable operation into the near THz-range. Methodologically, we have demonstrated the power of the synchrotron radiation based technique of SX-ARPES with its enhanced probing depth and sharp definition of the full three-dimensional **k** for the discovery of previously obscured properties of semiconductor heterostructures. Our results complement previous applications of SX-ARPES to buried oxide interfaces[7] and magnetic impurities in semiconductors[43] and topological insulators[44] which used elemental and chemical-state specificity of this technique achieved with resonant photoemission. In a broader perspective, our methodology arms the heterostructure growth technology with means to directly control the fundamental **k**-space parameters of the electronic structure, thereby delivering optimal transport and optical properties of the fabricated devices. Complementary to imaging of non-equilibrium electron motion in spatial coordinates,[45] we can envisage an extension of our experimental methodology to pump-probe experiments using X-ray free-electron laser (FELs) sources which will image the electron system evolution in **k**-space during transient processes in electronic devices.

## Methods

**Sample fabrication.** The GaN-HEMT heterostructures embedding a 2DEG were grown on *c*-oriented sapphire substrate in a SemiTeq STE3N MBE-system equipped with an ammonia ($NH_3$) nitrogen source. The buffer layer growth adopted the procedure described in Ref. 23. Prior to deposition, the substrate was annealed during 1 h and then nitrided for 40 min with 30 sccm $NH_3$ at 850°C. The following growth was carried out with 200 sccm $NH_3$. Deposition started with 20 nm AlN layer, grown at 1050°C. The following 200 nm AlN layer was grown at 1120°C with Ga used as a surfactant. Then a gradient junction to $Al_{0.43}Ga_{0.57}N$ with a thickness of ~250 nm was achieved by a gradual decrease of the substrate temperature down to $T$ = 830°C, followed by 140 nm of growth at constant $T$. Then a second gradient junction to $Al_{0.1}Ga_{0.9}N$ with a thickness of ~140 nm was formed by reducing $T$ of the Al effusion cell. Then a 500-nm GaN layer was grown. The growth was finished by deposition of a barrier layer consisting of 2 nm AlN and 1 nm $Al_{0.5}Ga_{0.5}N$ for the ARPES experiments, and 1 nm AlN and 15 nm $Al_{0.45}Ga_{0.55}N$ for measurements of transport properties. Hall effect characterization was carried out in magnetic fields up to 40 kG. The magnetic-field dependences were measured in both the standard Hall and van der Pauw geometries. These measurements were carried out at the Resource Center of Electrophysical Methods (Complex of NBICS-technologies of Kurchatov Institute).

**SX-ARPES experiments**. Raw SX-ARPES data were generated at the Swiss Light Source synchrotron radiation facility (Paul Scherrer Institute, Switzerland). The experiments have been carried out at the SX-ARPES endstation[46] of the ADRESS beamline[47], delivering high photon fluxes up to $10^{13}$ photons*s$^{-1}$*(0.01% BW)$^{-1}$. With the actual experimental geometry, *p*-polarized incident X-rays selected electron states symmetric relative to the $\overline{\Gamma M}$ and $\overline{\Gamma K}$ azimuths. The projection $K_x$ of the photoelectron momentum was directly measured through the emission angle along the analyser slit, $K_y$ is varied through tilt rotation of the sample, and $k_z$ through $hv$. The experiments were carried out at 12 K to quench the thermal effects reducing the coherent **k**-resolved spectral component at high photoelectron energies[48]. The combined (beamline and analyzer) energy resolution was ~150 meV, and the angular resolution of the PHOIBOS-150 analyzer was ~0.07°. The X-ray spot size in projection on the sample was 30x75 μm², which allowed us to control spatial homogeneity of our samples. Charging effects were not detected due to the small thickness of the AlGaN barrier layer.

**Electronic structure calculations.** First-principles calculations for bulk GaN have been carried out in the DFT framework as implemented in the pseudopotential *Quantum Espresso* code[49] using ultrasoft pseudopotentials. The electron exchange-correlation term was treated within the Generalized Gradient Approximation (GGA) using the Perdew–Burke–Ernzerhof functional. Self-consistent calculations for bulk GaN were performed with the plane wave cut-off energy 60 Ry and **k**-space sampling over a grid of 10x10x5 points in the BZ, and corrected with the 'scissors' operator to reproduce the experimental band gap. Calculations for the GaN-HEMT heterostructure used a 1x1 slab geometry with the supercell including 18 u.c. of GaN in the middle between 3 u.c of AlN at each end, Fig. 2g. Atomic coordinates in the supercell were relaxed, but imposing the lateral u.c. of bulk GaN until the Hellmann-Feynman forces on each atom were <30 meV/Å. The plane-wave cutoff energy was 25 Ry with a **k**-grid of 10x10x1 points. The Gaussian window for LDOS calculations was set to 0.05 eV.

**Transport measurements.** Raw transport data were generated at the Kurchatov Institute. To measure linear and nonlinear transport properties of the GaN-HEMT heterostructures, two types of test modules with low-resistance regrown ohmic contacts[50,51] were formed. Details on processing can be found in ref. 51. The first type modules were conventional transmission line measurement (TLM) modules with a channel width of 20 μm and channel lengths of 2.5 μm, 10 μm, 20 μm, and 40 μm (marked 1 in Fig. 5a). These modules were used to determine contact resistance, which was found to be 0.15 Ω*mm. Also *I-V* curves were measured at the smallest gaps (2.5 μm length channels, see Fig. 5b) of 24 such TLM modules (6 modules per each of four directions) in DC mode. The voltage sweep time (1 ms per point with a voltage step of 0.5 V) was chosen to be small enough to suppress sample heating effects as judged by the absence of hysteresis in the forward and backward voltage scans as well as repeatability of the *I-V* curves with a sweep time reduction. The second type 'resistor' modules (marked 2 in Fig. 5a) were arrays of 1 mm long and 20 μm wide stripes (25 stripes per module separated by 20 μm mesa isolation) with contact pads on each side. These modules had negligible contact resistance and were used for precise measurement of the 2DEG low-field conductivity in different directions.

**Data availability**. Derived data supporting the findings of this study are available from the corresponding author on request. The SX-ARPES data were processed using the custom package MATools available at https://www.psi.ch/sls/adress/manuals.

## Acknowledgements

We thank M. B. Tsetlin, V.G. Nazin, E. E. Krasovskii and G. Aeppli for fruitful discussions. M.- A.H. was supported by the Swiss Excellence Scholarship grant ESKAS-no.2015.0257. N.K.C. was partly supported by the Grant RFBR 16-07-01188.

## Author contributions

L.L.L. and V.N.S. conceived the SX-ARPES research at the Swiss Light Source. L.L.L., M.-A.H. and V.N.S. performed the SX-ARPES experiment supported by X.W., B.T. and T.S. L.L.L. and V.N.S. processed and interpreted the data. V.N.S. set theoretical description of the ARPES response. M.-A.H. performed the DFT calculations. M.L.Z., I.O.M., E.S.G. and V.G.V. conceived the GaN-HEMT project at the Kurchatov Institute. I.O.M. and I.S.E. grew the samples supported by M.L.Z. N.K.C. performed Hall characterization. E.S.G. implemented the 2DEG numerical model. I.O.M., I.A.C. and M.L.Z. fabricated the TLM modules and performed transport measurements. V.N.S. wrote the manuscript with contributions of L.L.L., V.G.V., E.S.G. and I.O.M. All authors discussed the results, interpretations, and scientific concepts.

## Additional information

**Supplementary Information** accompanies this paper at http://www.nature.com/naturecommunications

**Competing interests:** The authors declare no competing financial or non-financial interests.

# Figures

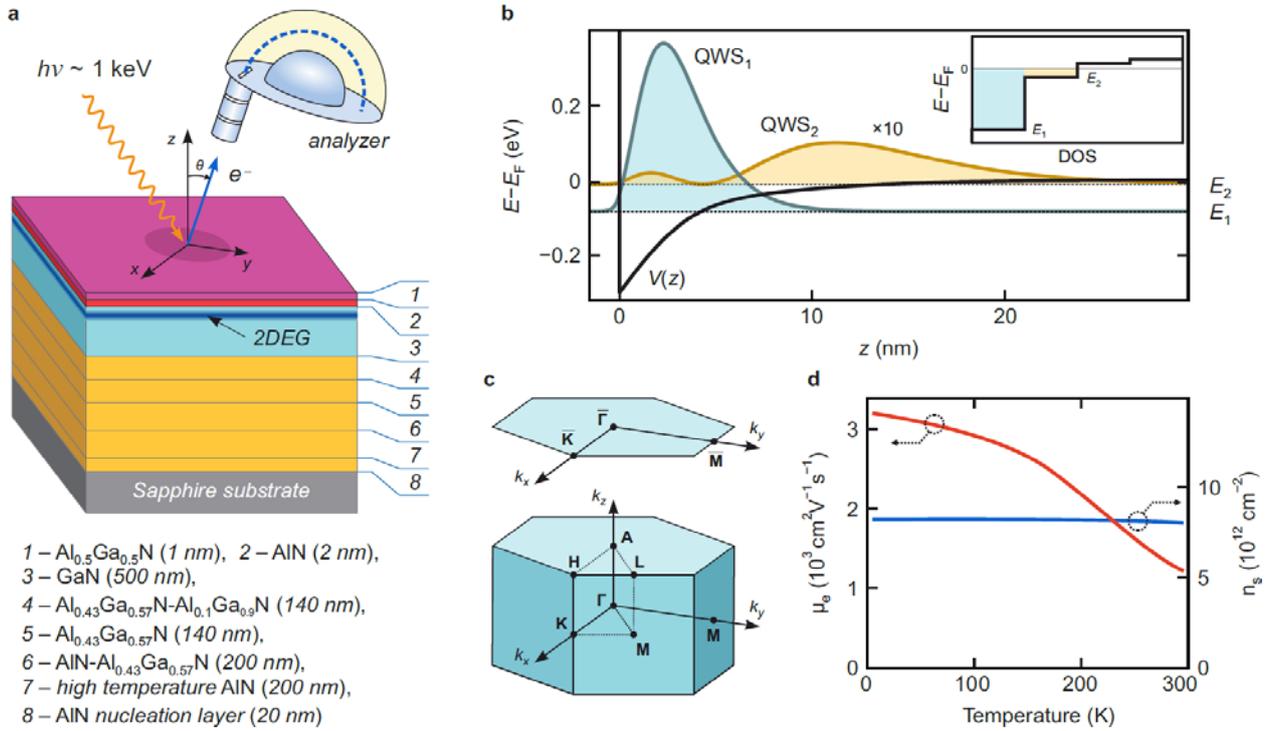

**Figure 1 | SX-ARPES experiment on the GaN-HEMT heterostructure. a**, Scheme of the epitaxial GaN-HEMT samples investigated by SX-ARPES. The photoelectron analyser detects the distribution $I_{PE}(E_k, \vartheta)$ of the photoelectron kinetic energy $E_k$ and emission angle $\vartheta$ which yield the binding energy $E_B$ and momentum **k** back in the sample (corrected for the photon momentum $p = h\nu/c$) to produce the sought-for electron dispersions $E(\mathbf{k})$. **b**, $T$-dependence of $n_s$ and $\mu_e$ obtained from Hall measurements. **c,** Sketch of the electronic structure based on self-consistent solution of the 1D Poisson-Schrodinger equation. The quasi-triangular 1D potential $V(z)$ confines two QWSs having different spatial localization of their electron density $n^i(z)$ (exaggerated by x10 for the QWS$_2$) centered at ~3 and ~12 nm for QWS$_1$ and QWS$_2$, respectively. The total three-dimensional DOS (*insert*) show steps characteristic of the 2D states. **d**, Bulk BZ of GaN and 2D one of the GaN/AlGaN heterostructure.

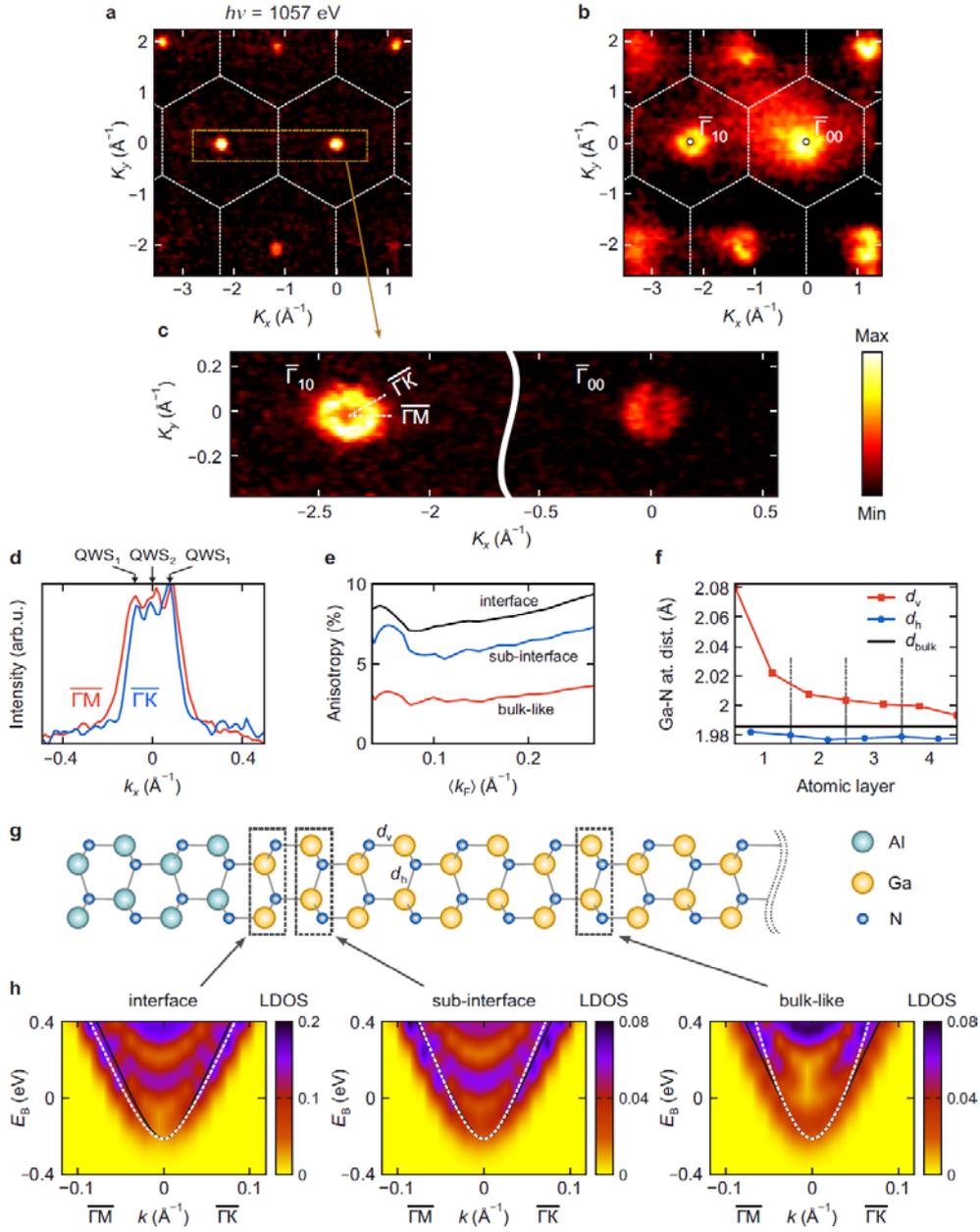

**Figure 2 | FS formed by the buried 2DEG. a**, Experimental FS formed by the 2DEG in comparison with **b**, iso-$E_B$ surface of the VB near the VBM. The FS appears as narrow electron pockets centered at the around the $\bar{\Gamma}$-points, consistent with the CBM-derived character of the 2DEG. Both VB and FS maps reflect the C6 symmetry of the GaN crystal lattice. **c**, FS along the $\bar{\Gamma}_{00}$-$\bar{\Gamma}_{10}$ line acquired with high energy and angle resolution. **d**, MDCs of the Fermi intensity around the $\bar{\Gamma}_{10}$-point (derived from the high-statistics data in Fig. 4) identifying the tiny $QWS_2$ and anisotropy of the $QWS_1$ between the $\bar{\Gamma M}$ and $\bar{\Gamma K}$ azimuths with $A_F \sim 12\%$. **e**, Calculated $A_F$ of $QWS_1$ between $\bar{\Gamma M}$ and $\bar{\Gamma K}$ as a function of band filling characterized by $\langle k_F \rangle$, for bulk GaN and for various heterostructure layers. **f**, relaxation of the Ga-N bond length as a function of depth, and **g**, u.c. used in the slab calculations. **h**, $k_{//}$-resolved LDOS for various heterostructure layers near the CBM with $E_F$ adjusted to the experimental $\langle k_F \rangle$ and superimposed with the corresponding bulk $E(\mathbf{k})$ along ΓM and ΓK (black dashed lines). The $QWS_1$ dispersion (white dashed in the bottom of the LDOS continuum) in the top GaN layers shows an asymmetry related to the interfacial atomic relaxation.

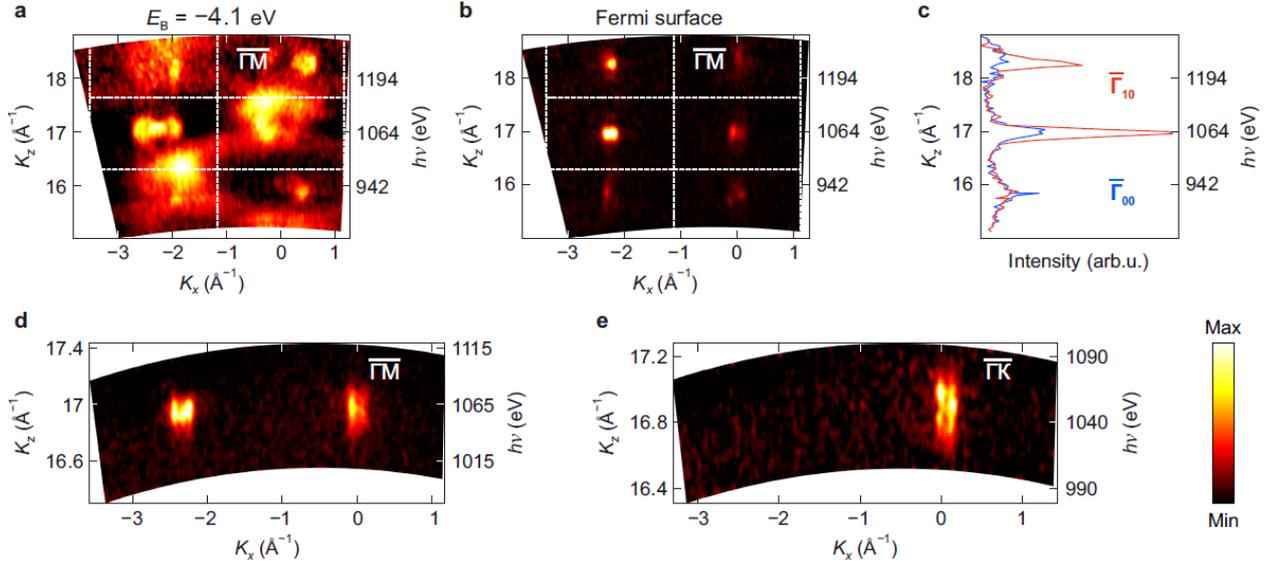

**Figure 3 | ARPES response of the 2DEG as a function of $K_z$ momentum.** ARPES intensity along the indicated azimuths is plotted as a function of $K_z$ rendered from $h\nu$ (the indicated $h\nu$ values correspond to the $\bar{\Gamma}$-point) **a**, Iso-$E_B$ map near the VBM, showing 3D contours of the VB states. **b**, FS formed by the QWSs. **c**, $E_F$-MDCs at the $\bar{\Gamma}_{00}$- and $\bar{\Gamma}_{10}$-points, showing periodically oscillating response of the QWSs. Its peaks located in the Γ-points evidence that the QWSs inherit their wavefunction from the CBM of parent bulk GaN. Zoom-in of the FS at the $\bar{\Gamma}_{00}$- and $\bar{\Gamma}_{10}$-points along the $\overline{\Gamma M}$ (**d**) and $\overline{\Gamma K}$ (**e**) measured at 1070 eV. The absence of its $K_z$ dispersion confirms the 2D character of the QWSs.

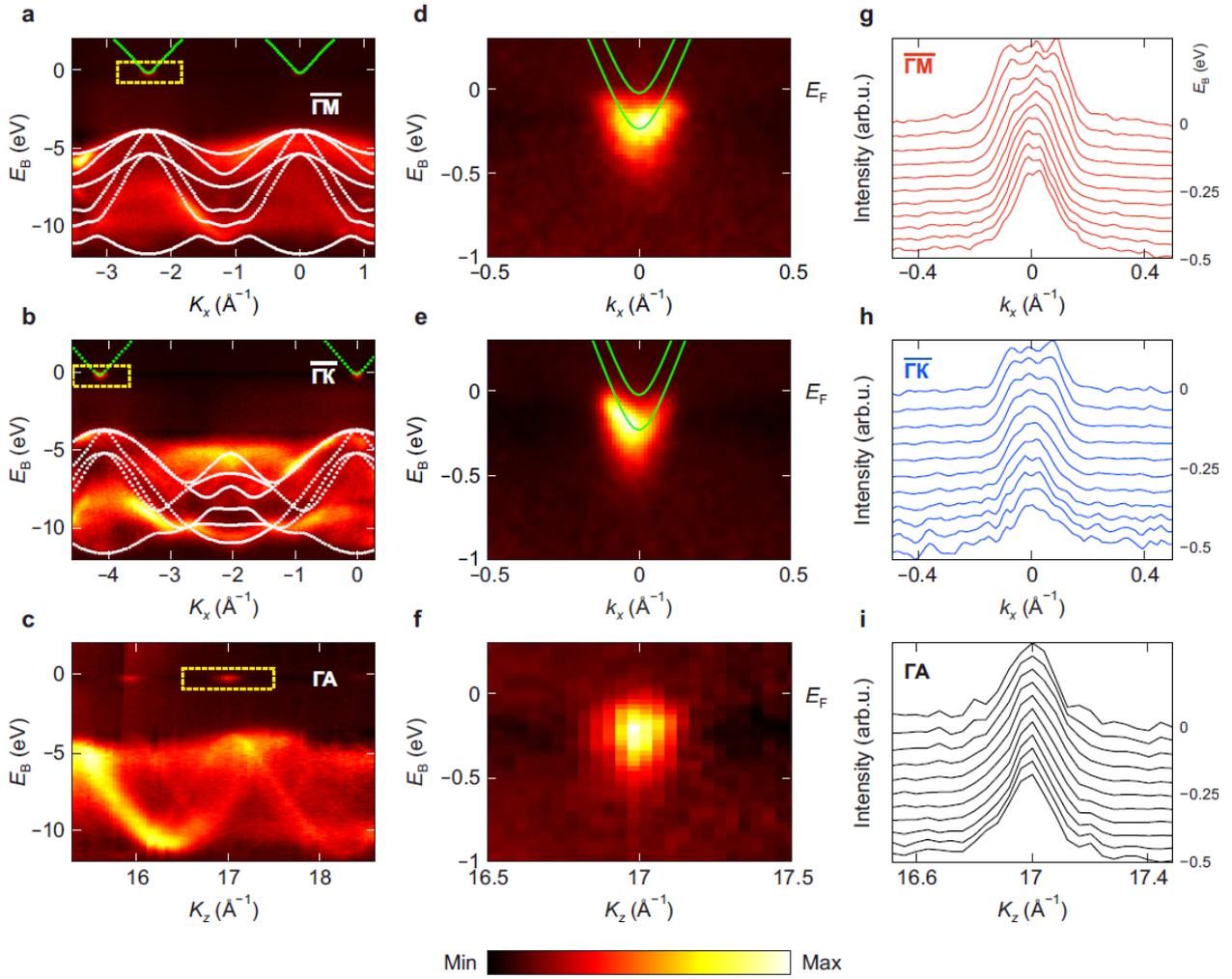

**Figure 4 | Band dispersions of the buried 2DEG. a-c**, Experimental band structure measured at $h\nu = 1066$ eV for the ΓK (**a**) and ΓM (**b**) directions of the bulk BZ (superimposed with $E(\mathbf{k})$ of bulk GaN, white dashed lines) and under variation of $h\nu$ for ΓA (**c**), cf. Fig. 3. The CBM-derived QWSs appear above the VB continuum. **d-f**, Zoom-in of the QWSs around the $\bar{\Gamma}_{10}$-point (green lines schematize their dispersions fitting the experimental $k_F$). **g-i**, (Normalized) MDCs around the $\bar{\Gamma}_{10}$-point for a series of $E_B$ through the QWS bandwidth. The difference between the $\overline{\Gamma M}$ and $\overline{\Gamma K}$ dispersions manifests planar anisotropy of the 2DEG, and the absence of $k_z$-dispersion confirms its 2D character.

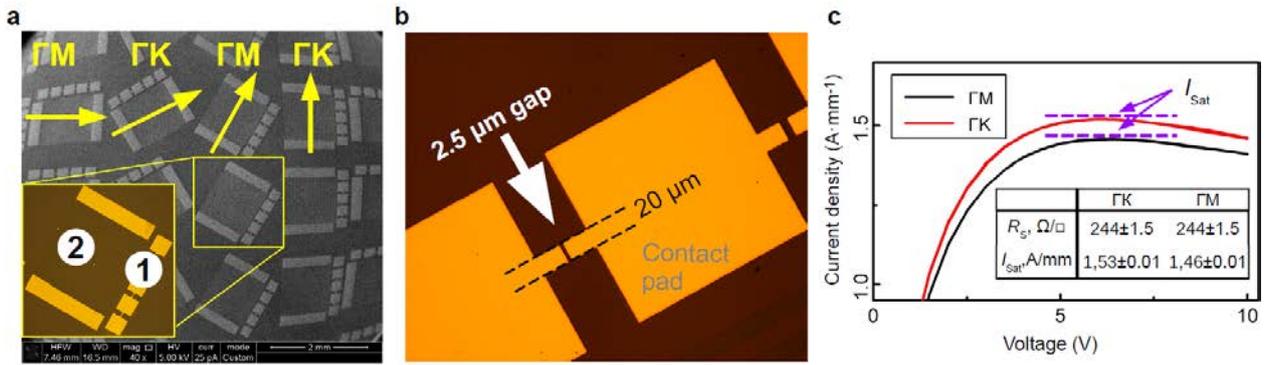

**Figure 5 | Electron transport measurements. a**, Test modules, oriented in four directions (scanning electron microscope images, slightly distorted due to large view area): TLM modules for the contact resistance measurements (marked 1 in the inset) and 'resistor' modules for $R_s$ determination (marked 2). The yellow arrows indicate the TLM azimuthal orientation. **b**, Region of the TLM modules with the channel length 2.5 μm used for the *I-V* measurements (optical microscope image). **c**, *I-V* characteristics for the $\overline{\Gamma M}$ and $\overline{\Gamma K}$ azimuths. The inset table summarizes the mean $R_s$ and $I_{sat}$ values for different azimuths. While $R_s$ is essentially isotropic, higher $I_{sat}$ for the $\overline{\Gamma K}$ azimuth in comparison with $\overline{\Gamma M}$ indicates lower $m^*$ of electrons moving along $\overline{\Gamma K}$.

# Supplementary Information

## k-space imaging of anisotropic 2D electron gas in GaN/GaAlN high-electron-mobility transistor heterostructures

by Lev et al.

# Supplementary Note 1
# 1D model of the GaN-HEMT heterostructure

To describe 2DEG localized at the GaN/AlGaN interface, we employ a simple unipolar Poisson-Schrodinger self-consistent model. The equilibrium electron density in 2DEG is given by

$$n(\mathbf{r}) = \int dE\, f(E)\rho(E,\mathbf{r}),\qquad(1)$$

where $f(E)$ is the Fermi-Dirac distribution and $\rho(E,\mathbf{r})$ is the local density of states in 2DEG that is determined by solving the Schrodinger equation.

The ideal system is translation invariant in the 2DEG plane ($\mathbf{r}_\parallel$), and the wavefunctions can be written in the factorized form

$$\Psi_{n,\mathbf{k}_\parallel}(\mathbf{r}) = e^{i\mathbf{k}_\parallel \mathbf{r}_\parallel}\psi_n(z),\quad \mathbf{r} \equiv (\mathbf{r}_\parallel, z),\quad \mathbf{k}_\parallel \equiv (k_x, k_y),\qquad(2)$$

where $\mathbf{k}_\parallel$ is the in-plane electron momentum, and $n$ is the quantum number that characterizes discrete quantum levels in the quantum well which is formed in the perpendicular direction ($z$). The wavefunction $\psi_n(z)$ satisfies the following Schrodinger equation:

$$\left[\frac{\hbar^2}{2}\left(\frac{k_x^2}{m_x(z)} + \frac{k_y^2}{m_y(z)} - \frac{d}{dz}\frac{1}{m_z(z)}\frac{d}{dz}\right) + V_{\text{CBM}}(z) + V(z)\right]\psi_n(z) = E_n(\mathbf{k}_\parallel)\psi_n(z),\qquad(3)$$

where $V_{\text{CBM}}(z)$ is the potential profile of the conduction band minimum, and $V(z)$ is the Hartree electrostatic potential energy due to other electrons that is to be determined self-consistently.

The problem can be greatly simplified if electron motion in the 2DEG plane can be decoupled from the out-of-plane. To this aim, we assume [1] that the in-plane effective masses $m_x(z)$ and $m_y(z)$ are independent of $z$ and equal to the corresponding bulk GaN values. This approximation is justified due to the very small penetration of $\psi_n(z)$ into the AlGaN barrier layer. Then (3) reduces to

$$\left[-\frac{\hbar^2}{2}\left(\frac{d}{dz}\frac{1}{m_z(z)}\frac{d}{dz}\right) + V_{\text{CBM}}(z) + V(z)\right]\psi_n(z) = E_n\psi_n(z),\qquad(4)$$

with the energy $E_n(\mathbf{k}_\parallel)$ given by

$$E_n(\mathbf{k}_\parallel) = E_n + \frac{\hbar^2}{2m_x}k_x^2 + \frac{\hbar^2}{2m_y}k_y^2.\qquad(5)$$

After the Schrodinger equation (4) is solved, the density of states in (1) can be computed:

$$\rho(E,\mathbf{r}) = \rho(E,z) = 2\sum_n |\psi_n(z)|^2 \int \frac{d^2\mathbf{k}_\parallel}{(2\pi)^2} \delta(E - E_n(\mathbf{k}_\parallel)) = \frac{\sqrt{m_x m_y}}{\pi\hbar^2} \sum_n |\psi_n(z)|^2 \theta(E - E_n), \quad (6)$$

where the factor 2 stands for the double spin degeneracy, $\theta(x)$ is the Heaviside step function, and the $\psi_n(z)$ are assumed to be normalized. Then the electron density is

$$n(\mathbf{r}) = n(z) = k_B T \frac{\sqrt{m_x m_y}}{\pi\hbar^2} \sum_n |\psi_n(z)|^2 \ln\left[1 + \exp\left(\frac{E_F - E_n}{k_B T}\right)\right], \quad (7)$$

where $T$ is the 2DEG temperature, and $E_F$ is the Fermi level.

The electrostatic potential $V(z) = -e\phi(z)$ is computed from the Poisson equation:

$$-\frac{d}{dz}\left(\varepsilon(z)\frac{d\phi(z)}{dz}\right) = 4\pi\rho(z) \quad (8)$$

with the full charge density $\rho(z)$ given by

$$\rho(z) = e\left[N_d^+(z) - n(z) + \sum_k \sigma_k \delta(z - z_k)\right]. \quad (9)$$

Here $N_d^+(z)$ is the concentration of ionized residual donors in GaN, and the last term describes the contribution of spontaneous and piezoelectric polarization charges with the surface density $\sigma_k$ localized at the heterostructure interface at positions $z_k$.

For the Poisson equation we assume the Dirichlet boundary condition at the outer AlN surface and the zero Neumann boundary condition deep into the GaN buffer layer:

$$\phi(0) = \phi_s, \quad \phi'(L) = E_z(L) = 0. \quad (10)$$

The unknown surface potential $\phi_s$ is adjusted so that the value of the total electron density in the quantum well

$$n_s = \int dz\, n(z) \quad (11)$$

matches the ARPES and Hall measurements results.

The Gummel-type self-consistent iterations between eqs. (4) and (8) have poor convergence. It can be significantly improved if (8) is turned into a non-linear Poisson equation with a potential-dependent full charge density functional $\rho[\phi(z)]$ that is obtained using the first-order perturbation theory for (4). Details of this procedure can be found in [2].

The exchange-correlation potential $V_{xc}(z)$ in the local density approximation can also be included in (4), however, it leads to just a few meV correction of the quantum levels [3] and is ignored here.

# Supplementary Note 2
# Slab calculations for the GaN/AlN interface

Atomic relaxation and electronic structure of the GaN/AlN interface was simulated with slab calculations (for details see Methods of the main text). The supercell used in these calculations, Fig. 1 (*top*), used a geometry of 1x1 u.c. of bulk GaN in the lateral direction and 18 u.c. in the perpendicular direction which were confined by 3 u.c of AlN on each side. The symmetric supercell geometry and thus potential distribution was essential to suppress spurious electron states. The atomic positions were relaxed under constraint of the lateral u.c. symmetry and basis parameters. They are characterized by significant variations of the *c*-oriented Ga-N bond length in the perpendicular direction (see Fig. 2 in the main text).

**k**-resolved layer density of states (LDOS) [4,5] calculated through the valence and conduction bands for different atomic planes in the GaN/AlN supercell is shown in Fig. 1 (*bottom*). It is compared with the corresponding bulk GaN band dispersions calculated for $k_z = 0$ and adjusted in energy to match the LDOS peaks at the VBM and CBM. In fact, the bulk is represented by already the 8th atomic layer in the middle of the supercell, which recaptures the bulk bond lengths and band structure of the bulk.

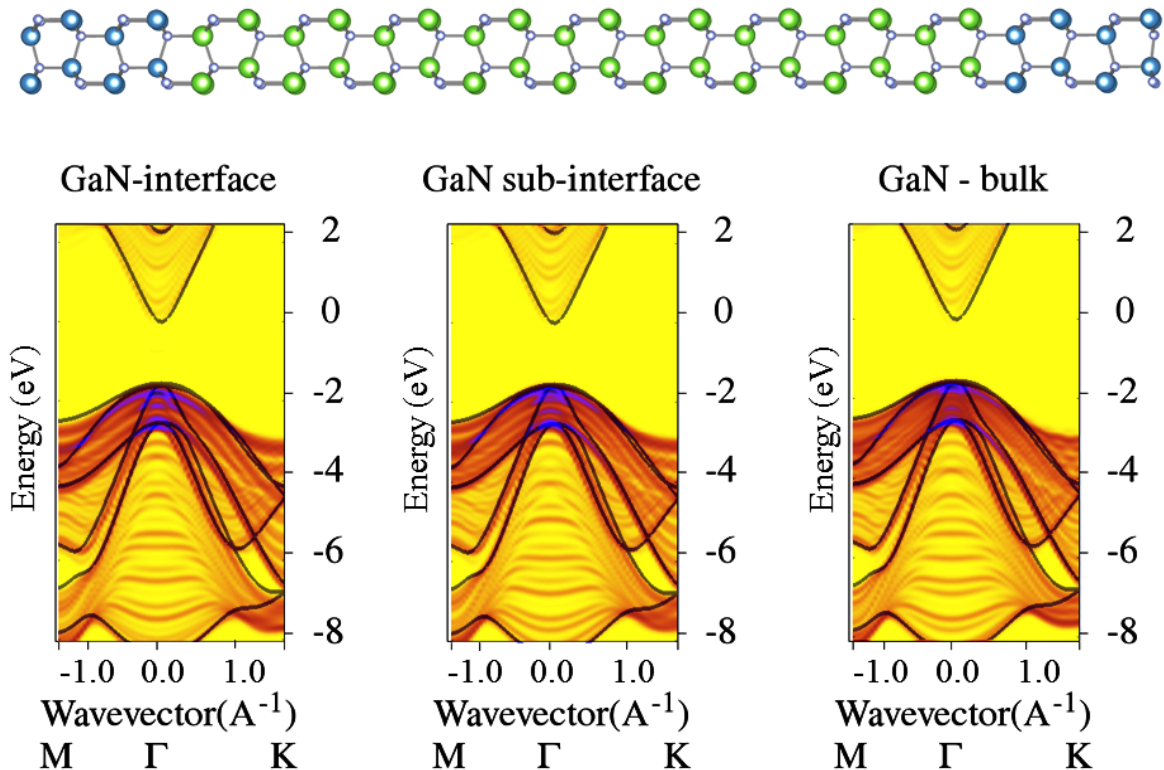

**Supplementary Figure 1.** (*top*) The supercell used in the calculations; (*bottom*) **k**-resolved LDOS for the GaN layer at the interface with AlN, for the 3-rd layer, and for the 8-th layer representing the GaN bulk. The superimposed solid curves are the bulk GaN bandstructure.

A zoom-in of the **k**-resolved LDOS from Fig. 1 in the CBM region is given in Fig. 2 of the main text. The LDOS outer contour corresponds to the $QWS_1$. Importantly, the $QWS_1$ dispersion in the interface layer demonstrates significant ΓM/ΓK asymmetry that progressively reduces towards the bulk. $k_F$ values for the $A_F$ factor quantifying the asymmetry (see the main text) were determined from parabolic fit of the $QWS_1$ dispersion in LDOS along ΓM and ΓK within ± 0.3 eV around the Fermi level. The evaluated $A_F$ as a function of band filling is presented in Fig. 2 of the main text.

Finally, we have performed the same self-consistent electronic structure calculations where the atomic coordinates were set to the bulk GaN lattice parameters without relaxation. The calculated LDOS in Fig. 2 demonstrates that in this case the interfacial ΓM/ΓK asymmetry reduces to its insignificant bulk values. This simulation supports our interpretation of the 2DEG anisotropy as resulting from relaxation of the interfacial atoms.

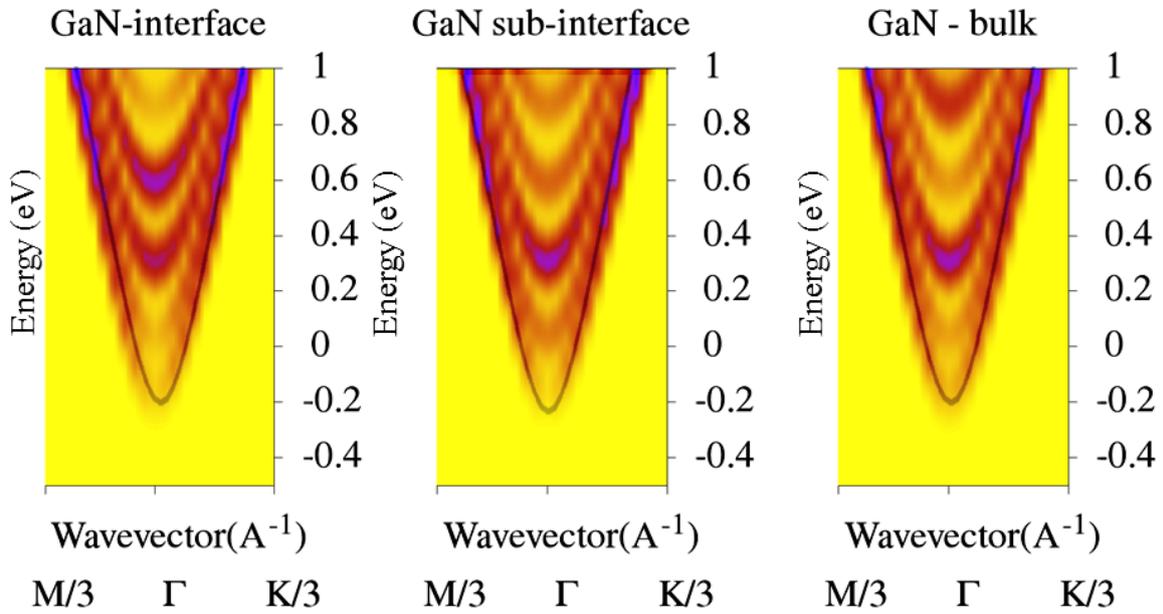

**Supplementary Figure 2.** The same LDOS in the CBM region as in Fig. 1 but calculated without relaxation of the interface atoms. The ΓM/ΓK asymmetry essentially disappears.

# Supplementary Note 3
# Crystal symmetry and electron conductivity

In this section we consider the crystal symmetry effect on linear (low-field) conductivity of crystals with anisotropic electronic structure.

The mathematics and discussions below can be in general form found in several textbooks (see, for example, [6]). However, implications of these principles to transport properties may be not obvious for non-specialists in transport properties of solid state systems. We have therefore compiled here these ideas in a focused form in order to save non-specialist readers from extensive literature search and calculations.

Applying external electric field $\vec{E}$ results in appearance of current with density $\vec{j} = \vec{j}(\vec{E})$. At low electric fields their relation is linear, and as such it has the following properties:

$$\vec{j}(\alpha \vec{E}) = \alpha \vec{j}(\vec{E}), \quad \text{(homogeneity)} \tag{1}$$

$$\vec{j}(\vec{E}_1 + \vec{E}_2) = \vec{j}(\vec{E}_1) + \vec{j}(\vec{E}_2), \quad \text{(additivity)}. \tag{2}$$

For any in-plane unit basis $(\vec{e}_x, \vec{e}_y)$ (we are interested in the fields and currents in the 2DEG plane) we have $\vec{E} = E_x \vec{e}_x + E_y \vec{e}_y$; using the properties (1), (2), we obtain:

$$\vec{j}(\vec{E}) = E_x \vec{j}(\vec{e}_x) + E_y \vec{j}(\vec{e}_y), \tag{3}$$

or, in the matrix form:

$$\begin{pmatrix} j_x \\ j_y \end{pmatrix} = \begin{pmatrix} \sigma_{xx} & \sigma_{xy} \\ \sigma_{yx} & \sigma_{yy} \end{pmatrix} \begin{pmatrix} E_x \\ E_y \end{pmatrix} = \hat{\sigma} \begin{pmatrix} E_x \\ E_y \end{pmatrix}, \text{ or } j_\alpha = \sigma_{\alpha\beta} E_\beta, \tag{4}$$

where $\hat{\sigma}$ is the conductivity tensor, $\alpha$ and $\beta$ run through the $x$ and $y$ coordinates, and the $\vec{j}(\vec{e}_x)$ and $\vec{j}(\vec{e}_y)$ vectors is decomposed in our 2D basis as $\vec{j}(\vec{e}_x) = \sigma_{xx} \vec{e}_x + \sigma_{xy} \vec{e}_y$, $\vec{j}(\vec{e}_y) = \sigma_{yx} \vec{e}_x + \sigma_{yy} \vec{e}_y$.

Now we will consider how the symmetry of a crystal restricts the form of tensors representing its physical properties. According to the Neumann's symmetry principle, if a crystal is invariant with respect to certain symmetry operations, all its physical properties must also be invariant with respect to this symmetry operations. In other words, the symmetry group of any physical property of a crystal must include the whole point symmetry group of the crystal. In particular, the latter means that if $\hat{A}$ is a symmetry transformation operator, then in the linear case $\hat{A}(\vec{j}(\vec{E})) = \vec{j}(\hat{A}(\vec{E}))$.

Let us consider two types of in-plane symmetry operations in Cartesian coordinate system: rotation and reflection. The matrix of rotation around the origin by angle $\theta$ is

$$\hat{R} = \begin{pmatrix} \cos\theta & \sin\theta \\ -\sin\theta & \cos\theta \end{pmatrix} \tag{5}$$

If system is symmetric with respect to rotation by angle $\theta$, the two following matrix products must be equal:

$$\hat{R}\hat{\sigma} = \begin{pmatrix} \sigma_{xx}\cos\theta + \sigma_{xy}\sin\theta & \sigma_{yx}\cos\theta + \sigma_{yy}\sin\theta \\ -\sigma_{xx}\sin\theta + \sigma_{xy}\cos\theta & -\sigma_{yx}\sin\theta + \sigma_{yy}\cos\theta \end{pmatrix} \quad (6)$$

$$\hat{\sigma}\hat{R} = \begin{pmatrix} \sigma_{xx}\cos\theta - \sigma_{xy}\sin\theta & \sigma_{xx}\sin\theta + \sigma_{xy}\cos\theta \\ \sigma_{yx}\cos\theta - \sigma_{yy}\sin\theta & \sigma_{yx}\sin\theta + \sigma_{yy}\cos\theta \end{pmatrix} \quad (7)$$

Equating the corresponding matrix components, we obtain

$$\begin{cases} \sigma_{xx}\sin\theta = \sigma_{yy}\sin\theta, \\ \sigma_{xy}\sin\theta = -\sigma_{yx}\sin\theta. \end{cases} \quad (8)$$

In particular, from (8) it follows that any $\theta: \sin\theta \neq 0$ immediately yields $\sigma_{xx} = \sigma_{yy}$.

Now consider a reflection, for simplicity, with respect to the abscissa axis. The corresponding transformation matrix

$$\hat{T} = \begin{pmatrix} 1 & 0 \\ 0 & -1 \end{pmatrix}, \quad (9)$$

renders the $(0, E_y)$ vector to $(0, -E_y)$ one, leaves the $(E_x, 0)$ vector unchanged. Equating the $\hat{\sigma}\hat{T}$ and $\hat{T}\hat{\sigma}$ products, one finds that $\sigma_{xx} = \sigma_{yy} = 0$, and therefore

$$\hat{\sigma} = \begin{pmatrix} \sigma_{xx} & 0 \\ 0 & \sigma_{yy} \end{pmatrix}. \quad (10)$$

As we see, symmetry with respect to rotation by an angle $\theta: \sin\theta \neq 0$, combined with the in-plane axial reflection symmetry, results in a diagonal conductivity tensor where the diagonal components are equal. I.e., the linear conductivity tensor is a scalar in this case. We should mention that the same considerations are valid for the case of (0001) plane of $C_6$ (hexagonal) point group crystals, (0001) plane of $C_{3v}$ point group crystals, (001) plane of cubic crystals, and other symmetrical systems fulfilling the above requirements.

# Supplementary References